\documentclass[12pt,a4paper]{article}

\usepackage{a4wide}
\usepackage{fancyhdr}
\usepackage{graphicx}
\usepackage{adjustbox}
\usepackage{caption}
\usepackage{subcaption}
\usepackage[ngerman, english]{babel}
\usepackage[utf8]{inputenc}
\usepackage[T1]{fontenc}
\usepackage{lmodern}
\usepackage{xcolor}
\usepackage[colorlinks,
citecolor={blue!70!black},
urlcolor={blue!70!black},
linkcolor={red!70!black},
hyperindex,breaklinks]{hyperref}
\usepackage{cite}
\usepackage{amsmath, amsthm, amssymb}
\usepackage{braket}
\usepackage{authblk}
\usepackage{tikz}
\usepackage{float}
\usepackage{indentfirst}

\title{Fast Pre-scramblers}

\author{Oleg Kaikov\thanks{Kaikov.Oleg@physik.uni-muenchen.de} $^{1,2}$\\
$^1${\small{\em Arnold Sommerfeld Center, Ludwig-Maximilians-Universit{\"a}t,
}}\\
{\small{\em Theresienstra{\ss}e 37, 80333 M{\"u}nchen, Germany
}}\\
$^2${\small{\em
Max-Planck-Institut f{\"u}r Physik,}}\\
{\small{\em F{\"o}hringer Ring 6, 80805 M{\"u}nchen, Germany
}}\\
}

\date{\small{\today}}

\begin{document}

\maketitle

\begin{abstract}
\noindent
We consider the process of diffusion or ``pre-scrambling'' of information in a quantum system. We define a measure for this spreading or ``pre-scrambling'' of the wavefunction in terms of a minimum probability threshold for the states in the system's Hilbert space. We illustrate our findings on the example of a prototype model with enhanced memory capacity. We conjecture:
\begin{itemize}
\item[(1)] The fastest pre-scramblers require a time logarithmic in the number of degrees of freedom.
\item[(2)] The investigated enhanced memory capacity model is a fast pre-scrambler.
\item[(3)] (Fast) pre-scrambling occurs not later than (fast) scrambling.
\item[(4)] Fast scramblers are fast pre-scramblers.
\item[(5)] Black holes are fast pre-scramblers.
\end{itemize}
\end{abstract}

\section{Introduction}

\subsection{Scrambling} \label{subsect_scr}

Every physical system carries information. Due to the time evolution of the system, the state of the information within it also evolves. Consider a quantum system of $K$ degrees of freedom, prepared in a pure state and evolved unitarily in time. Even though the state of the system remains pure throughout its unitary evolution, nevertheless, the system thermalizes after some time. This thermalization takes place in the following sense: although the time evolution is unitary, over time, the initial state becomes entangled with the other states in the Hilbert space of the system. The information initially contained only in the original state is distributed over the other states and is therefore mixed up. Thus, the system ``scrambles''~\cite{Hayden_2007,Sekino_2008} information. The characteristic time scale associated with this process is called the \emph{scrambling time} $t_s$~\cite{Hayden_2007,Sekino_2008}.

Before characterizing the scrambling time, we need to first address scrambling itself. Following~\cite{Hayden_2007,Sekino_2008,Susskind_2011}, which introduced the concept of scrambling, we call a system \emph{scrambled} if the information within it (i.e.\ the state of the system) is sufficiently distributed over the entire Hilbert space of the system with respect to some chosen measure.

Note that we keep the definition general on purpose: we do not limit the formulation to a specific measure of the uniformity of the state distribution. There exist multiple distinct such measures defined in the literature. Some of the first explicit definitions include the Haar measure in application to the choice of the mixing unitary transformation~\cite{Hayden_2007}, or to states (``Haar-scrambled'')~\cite{Sekino_2008}. Following~\cite{Page_1993}, for a model of $K$ qubits in~\cite{Sekino_2008}, the authors also define a system to be ``Page-scrambled'' when the entanglement entropy of any subsystem of $n<K/2$ qubits satisfies $S_n = n - O(\exp[2n-K])$. In general, a detailed picture of scrambling, especially for black holes, requires an understanding of scrambling at a microscopic level. In particular, the microscopic picture of scrambling, introduced in~\cite{Dvali_2013_1,Dvali_2014,
Dvali_2015_1,Dvali_2013_2,Dvali_2016_2} (see also further references therein), establishes the concept of ``memory modes'' (see the discussion below). Moreover, it discusses the concepts and the corresponding time-scales of ``one-particle entanglement''~\cite{Dvali_2013_2} and ``maximal entanglement''~\cite{Dvali_2016_2} for microscopic models of black holes.

However, in the present work we introduce a novel regime, which takes place on time-scales smaller than the various aforementioned scrambling time-scales. We thus establish a new pre-scrambling stage of a system's evolution, which requires quantification. In addition, we note that the results of the present paper are independent of the choice of a specific measure for scrambling.

What is the shortest possible scrambling time that a system can have? Following~\cite{Hayden_2007,Sekino_2008,Susskind_2011}, systems that scramble information in a time logarithmic in the number of degrees of freedom are the quickest scramblers. As in~\cite{Hayden_2007,Sekino_2008,Susskind_2011}, we refer to such systems as \emph{fast scramblers}. We broaden the above definition to also include the number of sites, modes, or generic quantum labels in a system, in addition to the number of degrees of freedom. Furthermore, to keep the definition general, we do not include any temperature dependence or any bound due to the latter, as is, for example, done in~\cite{Sekino_2008,Shenker_2014,Kitaev_2015,Maldacena_2016} among many other works. The results of the present paper apply in general, and require only that fast scramblers have a scrambling time that is logarithmic in the number of degrees of freedom $K$. The temperature dependent scrambling time for the model considered in section~\ref{sect_model} will be analyzed elsewhere~\cite{scr_to_appear}. In this paper we investigate the behavior of the above model at times preceding its scrambling time.

\subsection{Enhanced memory capacity}

Before considering the model of section~\ref{sect_model} we first need to introduce the concepts that are at its foundation. A sub-direction of the larger program of~\cite{Dvali_2018_1,Dvali_2019_1,Dvali_2013_1,Dvali_2014,
Dvali_2015_1,Dvali_2015_2,Dvali_2016_1,Dvali_2017,
Dvali_2018_2,Dvali_2018_3,Dvali_2019_2,Dvali_2019_3,
Dvali_2019_4,Dvali_2020_1,Dvali_2020_2, Dvali_2022_1, Dvali_2022_2, Dvali_2022_3, Dvali_2022_4, Dvali_2022_5} (see also further references therein) investigates general phenomena of systems that possess states with a high capacity to store information. As in~\cite{Dvali_2018_1,Dvali_2019_1,Dvali_2014,
Dvali_2015_1,Dvali_2015_2,Dvali_2016_1,Dvali_2017,
Dvali_2018_2,Dvali_2018_3,Dvali_2019_2,Dvali_2019_3,
Dvali_2019_4,Dvali_2020_1,Dvali_2020_2}, we refer to such systems as those with \emph{enhanced memory capacity}. One of the above phenomena is the effect of ``memory burden''~\cite{Dvali_2018_1,Dvali_2019_1}, which we briefly summarize below along with other terminology. For this, we largely rely on the recent detailed review included in~\cite{Dvali_2020_2}.

The degrees of freedom of a physical system are commonly described as quantum oscillators. The states of the system are then labeled by distinct sequences of the $K$ oscillators' occupation numbers $\ket{n_1, \dotsc, n_K}$. We refer to each such sequence as a \emph{memory pattern} which stores quantum information. The number of distinct patterns that can be stored within a microscopically narrow energy gap are a measure for the memory capacity of a system~\cite{Dvali_2018_2,Dvali_2019_2}. If the $\mathcal{N}$-many microstates $\ket{n_1, \dotsc, n_K}$ describing distinct patterns are degenerate in energy, they contribute to the microstate entropy $S = \ln (\mathcal{N})$. If a memory pattern contains a large amount of quantum information, it stabilizes the system in the state of enhanced memory capacity and slows down the system's evolution. Following~\cite{Dvali_2018_1,Dvali_2019_1}, we refer to this as the \emph{memory burden} effect.

A system can reach a state of enhanced memory capacity by the effect of \emph{assisted gaplessness}~\cite{Dvali_2019_2}. In its essence, assisted gaplessness occurs when a highly occupied \emph{master mode} interacts attractively with a set of \emph{memory modes}, lowering their energy gaps. The memory modes become effectively gapless and degenerate in energy, and can store large amounts of information. In their own turn, the memory modes backreact on the master mode via the memory burden effect~\cite{Dvali_2018_1,Dvali_2019_1} and slow down the change in its occupation number. As proposed by~\cite{Dvali_2018_1} and investigated in~\cite{Dvali_2020_2}, the memory burden effect can be avoided if the system possesses another set of modes $K'$ to which it can rewrite the information stored in the modes $K$.

\section{A prototype model} \label{sect_model}

We consider a specific prototype model of a system with enhanced memory capacity given in Eq.\ (34) of~\cite{Dvali_2020_2}, which is constructed in correspondence to the black hole's quantum $N$-portrait~\cite{Dvali_2013_1}. The model is given in Eq.\ (\ref{eqn_model}) with slight changes in notation for the sake of brevity. It contains two sets of bosonic memory modes $K$ and $K'$, with the corresponding creation and annihilation operators $\hat{a}_k^{\dagger}$, $\hat{a}_k$ and $\hat{a}_{k'}^{\dagger}$, $\hat{a}_{k'}$, respectively, for $k^{(\prime)}=1, \dotsc, K^{(\prime)}$. The operators $\hat{a}_{k^{(\prime)}}^{(\dagger)}$ obey the standard commutation relations (here and throughout we set $\hbar \equiv 1$)
\begin{equation} \label{eqn_comm_rel}
\left[ \hat{a}_{j^{(\prime)}}, \hat{a}^{\dagger}_{k^{(\prime)}} \right] = \delta_{j^{(\prime)}k^{(\prime)}}, \quad \left[ \hat{a}_{j^{(\prime)}}, \hat{a}_{k^{(\prime)}} \right] = 0, \quad \left[ \hat{a}^{\dagger}_{j^{(\prime)}}, \hat{a}^{\dagger}_{k^{(\prime)}} \right] = 0.
\end{equation}
The corresponding occupation number operators are given by $\hat{n}_{k^{(\prime)}}=\hat{a}_{k^{(\prime)}}^{\dagger}\hat{a}_{k^{(\prime)}}$ with eigenvalues $n_{k^{(\prime)}}$ and eigenstates $\ket{n_{k^{(\prime)}}}$. Additionally, the model contains two more bosonic modes $\hat{n}_a$ (the master mode) and $\hat{n}_b$, with creation and annihilation operators $\hat{a}^{\dagger}$, $\hat{a}$ and $\hat{b}^{\dagger}$, $\hat{b}$, respectively. These operators satisfy commutation relations analogous to Eq.\ (\ref{eqn_comm_rel}) and allow the exchange of occupation number between the modes $\hat{n}_a$ and $\hat{n}_b$. The couplings $C_m$ and $C_b$ parametrize the interaction strength among the memory modes $\hat{n}_{k^{(\prime)}}$, and between the modes $\hat{n}_a$ and $\hat{n}_b$, respectively.

The effective energy gaps of the $\hat{n}_k$ and $\hat{n}_{k^{(\prime)}}$ modes are
\begin{equation}
\varepsilon_k \equiv \varepsilon \left( 1- \dfrac{\hat{n}_a}{N} \right) \quad \text{and} \quad \varepsilon_{k'} \equiv \varepsilon \left( 1- \dfrac{\hat{n}_a}{N-\Delta} \right),
\end{equation}
correspondingly. Therefore, the modes $K$ and $K'$ become effectively gapless for $n_a=N$ and $n_a=N-\Delta$, respectively. In the beginning of the system's time evolution, $n_a=N$ and the $K$ modes are gapless, while the $K'$ modes are not. The $K$ modes initially contain all of the information in terms of excitations, whereas $K'$ modes are unoccupied. However, as $n_a$ diminishes from $N$ to $N-\Delta$, a state with $n_a=N-\Delta$, $n_b=\Delta$ and where the excitations have been transferred to the $K'$ modes, becomes energetically available.

We denote the total occupation number of the two memory sectors by
\begin{equation}
N_m \equiv \sum\limits_{k=1}^K n_k + \sum\limits_{k'=1}^{K'} n_{k'}.
\end{equation}
Note that it is conserved. Following~\cite{Dvali_2020_2}, we additionally truncate all memory modes to qubits. We denote the states of the system as
\begin{equation}
\ket{n_a, n_b, n_1, \dotsc, n_K, n_{1'}, \dotsc, n_{K'}} \equiv \ket{n_a} \otimes \ket{n_b} \bigotimes_{k=1}^K \ket{n_k} \bigotimes_{k'=1}^{K'} \ket{n_{k'}}.
\end{equation}
The system is evolved from an initial state
\begin{equation} \label{eqn_in_st}
\ket{in} = \ket{\underbrace{N}_{a}, \underbrace{0}_{b}, \underbrace{\overbrace{1, 1}^{=N_m}, 0, \dotsc, 0}_{K}, \underbrace{0, \dotsc, 0}_{K'}}
\end{equation}
where $\hat{n}_a$ and $\hat{n}_b$ have occupation numbers $N$ and $0$, respectively, and only the first $N_m$ modes of the initially gapless $K$ memory sector contain one particle each. Following~\cite{Dvali_2020_2}, we set the basic energy unit $e \equiv 1$. The Hamiltonian of the model is then
\begin{equation} \label{eqn_model}
\begin{aligned}
\hat{H} &= \varepsilon \left( 1 - \dfrac{\hat{n}_a}{N} \right) \sum\limits_{k=1}^{K} \hat{n}_k + \varepsilon \left( 1 - \dfrac{\hat{n}_a}{N-\Delta} \right) \sum\limits_{k'=1}^{K'} \hat{n}_{k'} + C_b \left( \hat{a}^{\dagger} \hat{b} + \text{H.c.} \right) + \\
&+ C_m \left\lbrace \sum\limits_{k=1}^{K} \sum\limits_{k'=1}^{K'} f_1(k,k') \left( \hat{a}_k^{\dagger} \hat{a}_{k'} + \text{H.c.} \right) + \sum\limits_{k=1}^{K} \sum\limits_{l=k+1}^{K} f_2(k,l) \left( \hat{a}_k^{\dagger} \hat{a}_l + \text{H.c.} \right) + \right. \\
&+ \left. \sum\limits_{k'=1}^{K'} \sum\limits_{l'=k'+1}^{K'} f_3(k',l') \left( \hat{a}_{k'}^{\dagger} \hat{a}_{l'} + \text{H.c.} \right) \right\rbrace,
\end{aligned}
\end{equation}
with
\begin{equation}
f_i(k,l) = \begin{cases}
		   F_i(k,l)-1, & F_i(k,l) < 0.5 \\
		   F_i(k,l),   & F_i(k,l) \geq 0.5
		   \end{cases}
\end{equation}
and
\begin{equation}
F_i(k,l) = \left( \sqrt{2}(k+\Delta k_i)^3 + \sqrt{7}(l+\Delta l_i)^5 \right) \text{ mod } 1,
\end{equation}
where
\begin{equation}
\Delta k_1 = \Delta k_2 = 1, \quad \Delta k_3 = K+1, \quad \Delta l_1 = \Delta l_3 = K+1, \quad \Delta l_2 = 1.
\end{equation}

For a spherically symmetric system, such as a black hole, the free energy gap $\varepsilon$ of the memory modes can be found by labeling the states by quantum numbers $(l,m)$ of spherical harmonics $Y_l^m(\theta,\phi)$~\cite{Dvali_2020_2}. From the number of independent solutions for one degree $l$ of a given $Y_l^m(\theta,\phi)$, the energy of the highest mode and hence the free gap of the memory modes can be estimated as~\cite{Dvali_2020_2}
\begin{equation}
\varepsilon = \sqrt{K}.
\end{equation}
Additionally, we require that in the vicinity of the initial state in Eq.\ (\ref{eqn_in_st}), the $K'$ memory sector is not gapless. This results in the constraint $|\varepsilon_{k'}| \gg 1/\sqrt{N_m}$~\cite{Dvali_2020_2} or, equivalently,
\begin{equation}
\Delta \gg \dfrac{N}{1+\sqrt{N_m}\sqrt{K}}.
\end{equation}
Due to the considerable duration of the numerical simulations, for the majority of parameter scans in section~\ref{subsect_indiv} we choose a value for $\Delta$ that does not satisfy this constraint. However, from the scan over $\Delta$ of the sought-after quantities, we are, firstly, able to extract a clear dependence over the majority of the domain, and secondly, we observe that one of the analyzed quantities acquires a dissimilar value only when $\Delta$ is in the immediate vicinity of $N$.
The requirement that the memory modes remain effectively gapless imposes the constraints on the couplings~\cite{Dvali_2020_2}
\begin{equation} \label{eqn_Cb_Cm_cond}
C_b \lesssim \dfrac{1}{\sqrt{N}}, \quad C_m \lesssim \dfrac{1}{\sqrt{N_m}\sqrt{K}}.
\end{equation}
Furthermore, as in~\cite{Dvali_2020_2}, we set
\begin{equation}
\begin{aligned}
K' &= K.
\end{aligned}
\end{equation}
This leaves a set of six free and constrained parameters
\begin{equation} \label{eqn_param}
N, K, N_m, \Delta, C_b, C_m.
\end{equation}
In terms of the black hole's quantum $N$-portrait~\cite{Dvali_2013_1}, for a black hole of entropy $S \gg 1$, the parameters are further constrained as $N=S, K=S, N_m=S/2$~\cite{Dvali_2020_2}. This scaling is considered in section~\ref{subsect_BH}.

\section{Two effects}

Based on the black hole's quantum $N$-portrait~\cite{Dvali_2013_1}, a microscopic framework to describe scrambling was established and subsequently developed in~\cite{Dvali_2013_2,Dvali_2013_3,Dvali_2014_1,Dvali_2015_1,Dvali_2015_2,
Dvali_2016_1,Dvali_2016_2,Dvali_2017_1,Dvali_2018_4,Kovtun_2020,
Kovtun_2022} among other works (see further references therein). However, as stated in section~\ref{subsect_scr}, our results are independent of the choice of a specific measure of the uniformity of the state distribution, which is necessary for a precise definition of scrambling. Instead, the only criterion we require a scrambled system to have, is for the system's state to be sufficiently distributed (with respect to some measure) over the \emph{entire} Hilbert space of the system. This is a property shared by all definitions of scrambling. Given this broad characterization, we expect two effects to occur not later than the instant in time the system becomes scrambled:
\begin{itemize}
\item[(1)] The system's wavefunction $\ket{\psi(t)}$, being initially in only one basis state $\ket{in}$, spreads over the entire basis $\{ \ket{v_i} \}$, given a minimum state probability threshold.
\item[(2)] The most probable basis state that the system can be found in, changes from $\ket{in}$ to some other basis state $\ket{v_i} \neq \ket{in}$.
\end{itemize}

Let us specify these effects and the times in which they occur more precisely. Namely, over some finite, non-zero time period, the wavefunction $\ket{\psi(t)}$ diffuses over the entire particle basis $\{ \ket{v_i} \}$, given a finite, fixed and sufficiently small precision $p$ to which the probabilities $|C_i|^2 = |\braket{\psi(t)|v_i}|^2$ of the individual basis states are rounded. For this arbitrarily small but fixed minimum state probability threshold $p < 1/\text{dim} \left( \{ \ket{v_i} \} \right)$, we denote the fraction of the basis that $\ket{\psi(t)}$ spreads over by $f$, and the time when $\ket{\psi(t)}$ spreads over the entire basis, i.e.\ the earliest time when $f=1$, by $t_f$. Concretely, we define $f$ as
\begin{equation}
f \equiv \dfrac{1}{\mathcal{N}} \sum\limits_{i=1}^{\mathcal{N}} H(|C_i|^2 - p)~, \quad
H(x) = \begin{cases}
	   0, & x \leq 0 \\
	   1, & x > 0
	   \end{cases},
\end{equation}
where $\mathcal{N} \equiv \text{dim} \left( \{ \ket{v_i} \} \right)$ is the number of basis states, i.e.\ the dimension of the Hilbert space of the system, and $H(x)$ is the left-continuous Heaviside step function. Clearly, for the model in Eq.\ (\ref{eqn_model}), $p$ has to be non-zero since the moment the system begins evolving, $\ket{\psi(t)}$ diffuses instantaneously over the entire basis, but with very small contributions of some states. However, at the time $t_f$, all of these probabilities increase above the given threshold of $p$.

Additionally, we denote by $t_c$ the time when the most probable state of the system changes from the initial basis state $\ket{in}$ to some other basis state $\ket{v_i} \neq \ket{in}$, i.e.\ the earliest time when $\underset{i}{\max} (|C_i|^2) \neq |C_{in}|^2$.

The above two effects are investigated numerically in section~\ref{sect_num} for the model in Eq.\ (\ref{eqn_model}). Our main goal is to understand how the two effects and their respective times $t_f$ and $t_c$ depend on the system's parameters in Eq.\ (\ref{eqn_param}). Note that the above definitions for the two effects of spreading involve quantitative measures of the quantum state's distribution in terms of probabilities. Due to this, our definitions differ somewhat to diffusion described in~\cite{Susskind_2011}. Furthermore, our definitions also differ from that of \emph{minimal scrambling} in~\cite{Dvali_2016_2}. Specifically, minimal scrambling for a system of $N$ qubits in~\cite{Dvali_2016_2} is defined as when the state of that system is a uniform superposition of $N$ basis states, each with $1/N$ probability, as opposed to a superposition of all $2^N$ basis states.

\section{Numerical results} \label{sect_num}

\begin{figure}[htbp]
  \centering
  \includegraphics[scale=0.75]{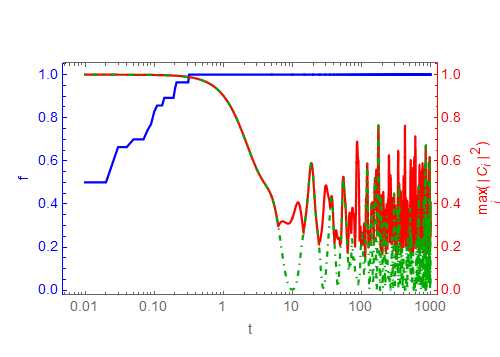}
  \caption{\small Left (blue): fraction of basis $f$ that the state $\ket{\psi(t)}$ is spread over. Basis includes $140$ states in total. Right (red): maximum contribution $|C_i|^2$ of any single basis state $\ket{v_i}$ with $i \in [1,140]$, and (green, dot-dashed): contribution of the initial state $|C_{in}|^2$. The time step is $10^{-2}$. Note that the time scale is logarithmic. In this example $t_f=0.32$ and $t_c=6.25$.}
  \label{fig_typ}
\end{figure}

The numerical simulations are performed with QuSpin version $0.3.4$~\cite{QuSpin_1,QuSpin_2,QuSpin_3}. Unless stated otherwise, and except for the parameter(s) varied, the parameter values we use are
\begin{equation}
p=5 \cdot 10^{-21}, \quad N=4, \quad K=4, \quad N_m=2, \quad \Delta=1, \quad C_b=0.1, \quad C_m=0.1.
\end{equation}
When varying the parameters, we restrict their possible ranges to respect the constraints in Eq.\ (\ref{eqn_Cb_Cm_cond}). By the value chosen for $p$ we imply that we round each individual $|C_i|^2$ to $20$ decimal places, and if the result is $>0$, we add its contribution to $f$. The reason that the precision $p$ needs be sufficiently small is that for $N$ and/or $K$ large enough, $t_f$ quickly diverges from growing logarithmically and rapidly increases. However, we expect that for an arbitrary but fixed range of values for $N$ and/or $K$ there exists a sufficiently small precision $p$ such that $t_f$ behaves logarithmically on the given parameter range. Figure~\ref{fig_typ} depicts an exemplary plot of the two effects for the chosen parameter values.

In several simulations we observe that after reaching $1$, the value of $f$ can decrease back below $1$ for a period of time, but then increases to $1$ and remains at this value. However, in most simulations, once $f=1$ is reached, it tends to stay at $1$. The maximal contribution $\underset{i}{\max} (|C_i|^2)$ can also come back to being equal to $|C_{in}|^2$, but then changes again to that of some other state. We also observe that the time $t_f$ does not always precede $t_c$; this is clear as the value of $f$ strongly depends on the chosen precision $p$.

We expect large values of $N$, $K$, $N_m$ (i.e.\ large systems with large Hilbert spaces) and $\Delta$ (i.e.\ large effective energy gap difference), and small values of $C_b$, $C_m$ (i.e.\ weak couplings) to slow down the spread of the wavefunction $\ket{\psi(t)}$. The goal of the numerical simulations is first, to verify and second, to quantify this behavior. Moreover, we expect the diffusion of $\ket{\psi(t)}$ as defined above in terms of the fraction $f$ of the basis to be a crucial pre-requisite for scrambling. This stems from a simple fact that before the system becomes scrambled over the entire basis with respect to any measure, its state has to firstly spread over the entire basis with some small non-zero contribution of every basis state. Furthermore, we expect there to be no ``diffusers'', ``spreaders'', ``distributors'' or ``pre-scramblers'' that are faster than the model in Eq.\ (\ref{eqn_model}), since in this system every mode of each sector is directly coupled to every other mode in that sector. It remains to investigate, how fast this ``fast pre-scrambler'' is.

\subsection{Individual parameter scaling} \label{subsect_indiv}

\begin{figure}[htbp]
    \centering
    \begin{subfigure}[b]{0.45\textwidth}
    	\centering
        \includegraphics[width=\textwidth]{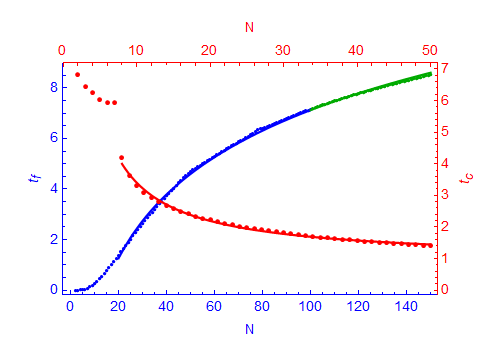}
        \caption{\footnotesize $t_f \sim \ln(N)$ for $N \geq 20$, \\ $t_c \sim N^{-0.966 \pm 0.044}$ for $N \geq 8$.}
    \end{subfigure}
    \begin{subfigure}[b]{0.45\textwidth}
        \includegraphics[width=\textwidth]{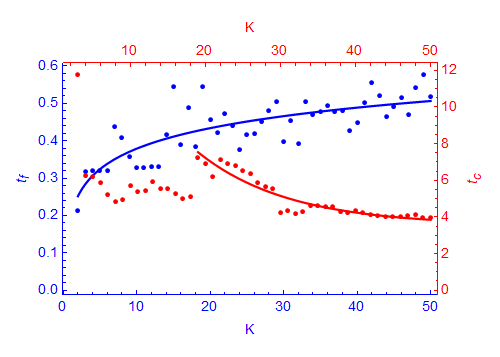}
        \centering
        \caption{\footnotesize $t_f \sim \ln(K)$, \\ $t_c \sim ~ \mathrm{e}^{(-0.083 \pm 0.019) K}$ for $K \geq 19$.}
    \end{subfigure}
    \begin{subfigure}[b]{0.45\textwidth}
        \includegraphics[width=\textwidth]{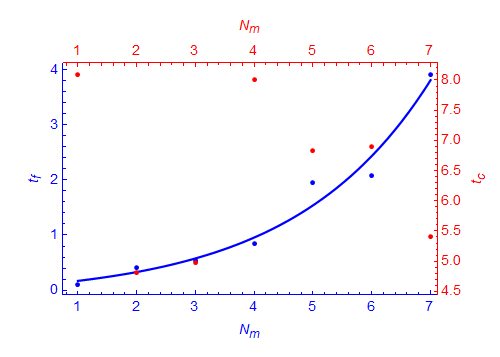}
        \caption{\footnotesize $t_f \sim ~ \mathrm{e}^{(0.43 \pm 0.13) N_m}$, \\ $t_c$ scaling unknown.}
    \end{subfigure}
    \begin{subfigure}[b]{0.45\textwidth}
        \includegraphics[width=\textwidth]{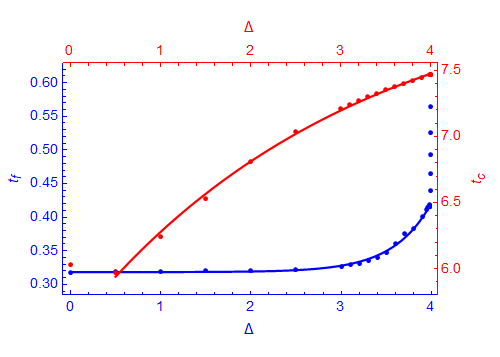}
        \caption{\footnotesize $t_f \sim \mathrm{e}^{(2.373 \pm 0.087) \Delta}$ for $\Delta \leq 3.986$, \\ $t_c \sim \mathrm{e}^{(-0.323 \pm 0.019) \Delta}$ for $\Delta \geq 0.5$.}
    \end{subfigure}
    \begin{subfigure}[b]{0.45\textwidth}
        \includegraphics[width=\textwidth]{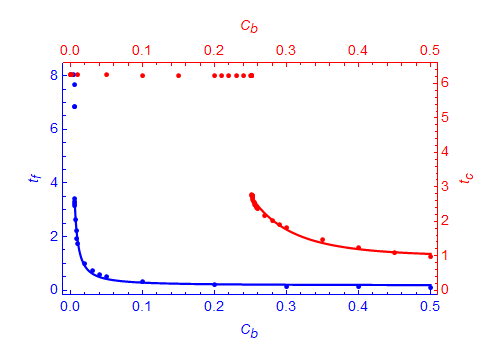}
        \caption{\footnotesize $t_f \sim C_b^{-1.177 \pm 0.059}$ for $C_b \geq 0.0057$, \\ $t_c \sim C_b^{-4.41 \pm 0.48}$ for $C_b \geq 0.25207$.}
    \end{subfigure}
    \begin{subfigure}[b]{0.45\textwidth}
        \includegraphics[width=\textwidth]{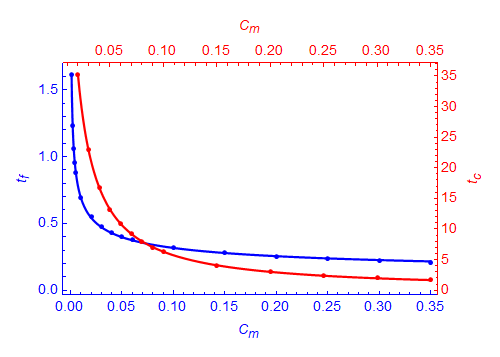}
        \caption{\footnotesize $t_f \sim C_m^{-0.3847 \pm 0.0051}$, \\ $t_c \sim C_m^{-1.0734 \pm 0.0044}$.}
    \end{subfigure}
    \caption{\small The individual subfigures show the scaling of $t_f$ and $t_c$ with the individual parameters $N$, $K$, $N_m$, $\Delta$, $C_b$, $C_m$. Note the different parameter ranges for the top and bottom, and left and right axes. The parameters altered are: (a) for only the $t_f$ scan: $p=5 \cdot 10^{-41}$, additionally, $N$ is varied up to $150$ for a better fit; the permitted range is $N \leq 100$. The points corresponding to $N \in [101,150]$ and the respective fit are marked green. (c) for both $t_f$ and $t_c$ scans: $K=8$.}
    \label{fig_indiv}
\end{figure}

We first vary the parameters of the system individually. The results are shown in Fig.\ \ref{fig_indiv} with the respective best obtained fit functions. The corresponding fit parameters are listed in Tab.\ \ref{tab} in the Appendix. We observe that indeed, large values of $N$, $K$, $N_m$, $\Delta$ and small values of $C_b$, $C_m$ slow down the diffusion in terms of $t_f$. Moreover, we obtain new quantitative information from the dependence of $t_f$ and $t_c$ on the model parameters. However, for the dependence of $t_c$ on $N_m$ no behavior could be extracted due to irregularity of the data points. Furthermore, note that for the dependence of $t_f$ on $N_m$ the standard errors of the free fit-model parameters are comparable in magnitude with the values of the parameters themselves, and that the adjusted coefficient of determination $\overline{R}^2$ and the adjusted root-mean-square error are the lowest and highest, respectively, when compared to all other fits. We therefore can not consider this to be a good fit to the data. Unfortunately, due to numerical limitations, we weren't able to obtain more data in a reasonable amount of time.

However, for higher values of $N_m$ we expect $t_f$ to increase logarithmically with $N_m$ for the following two reasons. Firstly, note that $t_f$ doesn't increase logarithmically also for small values of $N$. The logarithmic behavior becomes evident only for higher values of $N$. Secondly, there are only three parameters in the model, namely $N$, $K$ and $N_m$, that affect the dimension of the Hilbert space. We expect the size of the Hilbert space to be the main factor that controls the speed at which the wavefunction spreads, and hence also the behavior of $t_f$. We therefore expect $t_f$ to depend on $N$, $K$ and $N_m$ in identical manner.

Nevertheless, we are able to extract the scaling of $t_f$ and $t_c$ with respect to all other parameters. Our main finding is that
\begin{equation}
t_f \sim \ln(K).
\end{equation}
That is, the time $t_f$ for the initial state of the system to spread over the entire basis $\{ \ket{v_i} \}$ with $|C_i|^2 > p \; \forall i \in [1,\mathcal{N}]$ for a fixed $p$ scales logarithmically with the number of modes $K$. Since the model we are investigating possesses all-to-all couplings between the modes of each sector, we expect the above logarithmic scaling of the wavefunction's ``diffusion'', ``spreading'', ``distribution'' or ``pre-scrambling'' to be the fastest possible for any physical system.
\bigskip
\\
\makebox[\textwidth][c]{%
\begin{minipage}{0.8\textwidth}
We therefore define a system to be a \emph{fast pre-scrambler}, if for an arbitrarily large but fixed range of number of degrees of freedom $K$ there exists a sufficiently small minimum state probability threshold $p<1/\text{dim}(\{ \ket{v_i} \})$, such that the state of the system $\ket{\psi(t)}$, initially in one basis state $\ket{in}$, spreads over the entire Hilbert space $\{ \ket{v_i} \}$ into a superposition of all $\text{dim}(\{ \ket{v_i} \})$ basis states each with a probability $|C_i|^2 = |\braket{\psi(t)|v_i}|^2 > p$, in a time logarithmic in $K$.
\end{minipage}}
\bigskip
\\
There are two reasons why we name this effect ``pre-scrambling''. Firstly, we want to distinguish it from diffusion in~\cite{Susskind_2011} and minimal scrambling in~\cite{Dvali_2016_2}. Secondly, since the probability threshold $p$ can be set arbitrarily small, this type of the state's spreading will precede scrambling where the state is required to be sufficiently distributed with respect to some measure. One could, of course, define a scrambling measure identical to our pre-scrambling measure. However, in that case there is a possibility that while the state does spread over the entire basis to some extent, the most probable state of the system is still its initial state. In other words, this can occur if $t_f < t_c$. In this case it is questionable whether or not we can call the state of the system truly scrambled.

\subsection{Application to black holes} \label{subsect_BH}

Following the black hole's quantum $N$-portrait~\cite{Dvali_2013_1}, the parameters are constrained as mentioned in section~\ref{sect_model}. Unfortunately, an extensive scan with the corresponding scaling of parameters is very time consuming. However, we can still argue what we expect the behavior of $t_f$ and $t_c$ to be in this regime. Firstly, as was discussed previously, we expect $t_f$ to depend logarithmically on $N$, $K$ and $N_m$. Secondly, we performed scans of $t_f$ and $t_c$ for $N=K \leq 15$ with $N_m$ fixed constant and equal to $2$, with precisions $p=5 \cdot 10^{-41}$ and $p=5 \cdot 10^{-21}$, respectively. From these scans we observe that both $t_f$ and $t_c$ each scale as their individual $N$ scans.

Moreover, as is argued in~\cite{Hayden_2007,Sekino_2008}, black holes are fast scramblers, and are expected to scramble in a time logarithmic in the number of degrees of freedom. Based on the discussion above, we expect pre-scrambling to occur not later than scrambling, irrespective of the specific measure used to define the latter. The model in Eq.\ (\ref{eqn_model}) possesses direct all-to-all couplings between the modes of every sector. We therefore expect it to have the fastest pre-scrambling time possible in terms of the number of degrees of freedom. We have argued above that for this model pre-scrambling should occur in a time logarithmic in the number of degrees of freedom when varying $N$, $K$ and $N_m$ individually. From the arguments above it is then natural to expect the model in question to pre-scramble in a time logarithmic in $N$ for the above black hole parameter scaling regime.

\section{Conjectures}

Summarizing the results of the investigation above, we arrive at the following conjectures:
\bigskip
\\
\makebox[\textwidth][c]{%
\begin{minipage}{0.8\textwidth}
\begin{itemize}
\item[(1)] The fastest pre-scramblers spread the initial state of the system over its entire Hilbert space in a time logarithmic in the number of degrees of freedom, given a minimum state probability threshold.
\item[(2)] The enhanced memory capacity model in Eq.\ (\ref{eqn_model}) is a fast pre-scrambler.
\item[(3)] The time of (fast) pre-scrambling is $\leq$ than that of (fast) scrambling, irrespective of the choice of a specific measure of the uniformity of the state distribution for the latter.
\item[(4)] Consequently, fast scramblers are fast pre-scramblers.
\item[(5)] In particular, black holes are fast pre-scramblers.
\end{itemize}
\end{minipage}}
\bigskip
\\
Note, however, that we do \emph{not} claim any of the following: we do not claim that all fast pre-scramblers are fast scramblers, nor do we claim that all fast pre-scramblers have all-to-all couplings between the degrees of freedom. Furthermore, the above are conjectures and should by no means be regarded as concrete proofs.

\section{Conclusion}

In addition to formulating the above conjectures, we would like to conclude by several complementary comments. First, note that the diffusion of the wavefunction as measured by $t_f$ possesses an at least exponential dependence on the energy gap difference $\Delta$, and for $\Delta$ close to $N$ increases even more rapidly than for lower values of $\Delta$. Second, we find it encouraging and fascinating that even the simple prototype model of a system with enhanced memory capacity in Eq.\ (\ref{eqn_model}) exhibits properties that we expect black holes to possess. As an additional remark on the prototype model in Eq.\ (\ref{eqn_model}), note that the all-to-all direct coupling of modes is in full agreement with the description of a black hole as a fast scrambler in~\cite{Susskind_2011}. Third, we allow ourselves to speculate that the Sachdev-Ye-Kitaev model~\cite{Sachdev_1993,Kitaev_2015} is a fast pre-scrambler, since it possesses all-to-all direct couplings and was shown to be a fast scrambler.

We also note that from the numerical data we are able to observe that a large amount of quantum information stored in the system's state slows down the process of the system's pre-scrambling via the memory burden effect.

As an outlook, let us note that the temperature dependent scrambling as discussed in~\cite{Sekino_2008,Shenker_2014,Kitaev_2015,Maldacena_2016} will be investigated for the model in Eq.\ (\ref{eqn_model}) in a future work~\cite{scr_to_appear}. Finally, the scrambling for the model in Eq.\ (\ref{eqn_model}) in relation to the eigenstate thermalization hypothesis~\cite{Deutsch_1991,Srednicki_1994} will also be considered in a future work~\cite{eth_to_appear}.

\section*{Acknowledgments}

We thank Gia Dvali for insightful discussions and valuable comments.

\begin{appendix}

\section*{Appendix}

\begin{table}[H]
\centering
\begin{adjustbox}{width=\textwidth}
\begin{tabular}{l|l|l|l|l|l|l|l|l}
$t_f$ & $\overline{R}^2$ & $\text{RMSE}$ & $a$ & $\sigma_{a}$ & $b$ & $\sigma_{b}$ & $c$ & $\sigma_{c}$ \\
\hline
$a + b\ln(N)$ & $9.9991 \cdot 10^{-1}$ & $6.00 \cdot 10^{-2}$ & $-9.645$ & $4.3 \cdot 10^{-2}$ & $3.6420$ & $9.8 \cdot 10^{-3}$ & $\text{n/a}$ & $\text{n/a}$ \\
$a + b\ln(K)$ & $9.88 \cdot 10^{-1}$ & $4.80 \cdot 10^{-2}$ & $1.97 \cdot 10^{-1}$ & $2.8 \cdot 10^{-2}$ & $7.92 \cdot 10^{-2}$ & $8.8 \cdot 10^{-3}$ & $\text{n/a}$ & $\text{n/a}$ \\
$a + b\exp(cN_m)$ & $9.77 \cdot 10^{-1}$ & $2.86 \cdot 10^{-1}$ & $-1.3 \cdot 10^{-1}$ & $4.3 \cdot 10^{-1}$ & $1.9 \cdot 10^{-1}$ & $1.9 \cdot 10^{-1}$ & $4.3 \cdot 10^{-1}$ & $1.3 \cdot 10^{-1}$ \\
$a + b\exp(c\Delta)$ & $9.9996 \cdot 10^{-1}$ & $2.44 \cdot 10^{-3}$ & $3.183 \cdot 10^{-1}$ & $1.0 \cdot 10^{-3}$ & $7.7 \cdot 10^{-6}$ & $2.7 \cdot 10^{-6}$ & $2.373$ & $8.7 \cdot 10^{-2}$ \\
$a + b C_b^c$ & $9.98 \cdot 10^{-1}$ & $7.81 \cdot 10^{-2}$ & $1.76 \cdot 10^{-1}$ & $3.8 \cdot 10^{-2}$ & $7.2 \cdot 10^{-3}$ & $2.2 \cdot 10^{-3}$ & $-1.177$ & $5.9 \cdot 10^{-2}$ \\
$a + b C_m^c$ & $9.9993 \cdot 10^{-1}$ & $6.10 \cdot 10^{-3}$ & $5.34 \cdot 10^{-2}$ & $7.5 \cdot 10^{-3}$ & $1.086 \cdot 10^{-1}$ & $4.1 \cdot 10^{-3}$ & $-3.847 \cdot 10^{-1}$ & $5.1 \cdot 10^{-3}$

\bigskip
\bigskip \\

$t_c$ & $\overline{R}^2$ & $\text{RMSE}$ & $a$ & $\sigma_{a}$ & $b$ & $\sigma_{b}$ & $c$ & $\sigma_{c}$ \\
\hline
$a + b N^c$ & $9.996 \cdot 10^{-1}$ & $4.52 \cdot 10^{-2}$ & $9.25 \cdot 10^{-1}$ & $5.9 \cdot 10^{-2}$ & $22.9$ & $1.9$ & $-9.66 \cdot 10^{-1}$ & $4.4 \cdot 10^{-2}$ \\
$a + b\exp(c K)$ & $9.94 \cdot 10^{-1}$ & $3.97 \cdot 10^{-1}$ & $3.54$ & $3.2 \cdot 10^{-1}$ & $19.2$ & $6.3$ & $-8.3 \cdot 10^{-2}$ & $1.9 \cdot 10^{-2}$ \\
$a + b\exp(c\Delta)$ & $9.99994 \cdot 10^{-1}$ & $1.69 \cdot 10^{-2}$ & $8.205$ & $7.2 \cdot 10^{-2}$ & $-2.662$ & $5.3 \cdot 10^{-2}$ & $-3.23 \cdot 10^{-1}$ & $1.9 \cdot 10^{-2}$ \\
$a + b C_b^c$ & $9.991 \cdot 10^{-1}$ & $7.01 \cdot 10^{-2}$ & $9.72 \cdot 10^{-1}$ & $7.2 \cdot 10^{-2}$ & $3.9 \cdot 10^{-3}$ & $2.7 \cdot 10^{-3}$ & $-4.41$ & $4.8 \cdot 10^{-1}$ \\
$a + b C_m^c$ & $9.9998 \cdot 10^{-1}$ & $5.55 \cdot 10^{-2}$ & $3.0 \cdot 10^{-2}$ & $4.7 \cdot 10^{-2}$ & $5.288 \cdot 10^{-1}$ & $9.4 \cdot 10^{-3}$ & $-1.0734$ & $4.4 \cdot 10^{-3}$
\end{tabular}
\end{adjustbox}
\caption{Tables of best obtained fit functions for $t_f$ and $t_c$ on the top and bottom, respectively. The corresponding coefficients of determination $\overline{R}^2$ and unbiased root-mean-square errors (RMSE) are both adjusted for the number of fit-model parameters. The values of the free fit-model parameters are given with the corresponding standard errors. No dependence of $t_c$ on $N_m$ could be extracted from the corresponding data points.}
\label{tab}
\end{table}

\end{appendix}

\end{document}